\documentstyle[12pt]{article}
\font\titlefont=cmbx10 scaled \magstep4

\begin{document}
\input{epsf}

\begin{flushright}
\vspace*{-2cm}
gr-qc/9711030 \\
TUTP-97-11 \\  November 8, 1997
\vspace*{1cm}
\end{flushright}

\begin{center}
{\titlefont QUANTUM INEQUALITIES \\
\vspace*{0.1in}
AND SINGULAR NEGATIVE \\ 
\vspace*{0.15 in}
ENERGY DENSITIES}\\
\vskip .7in
L.H. Ford\footnote{email: ford@cosmos2.phy.tufts.edu},
Michael J. Pfenning\footnote{email: mitchel@cosmos2.phy.tufts.edu}, and 
Thomas A. Roman\footnote{Permanent address: Department of Physics and Earth
Sciences, Central Connecticut State University, New Britain, CT 06050 \\
email: roman@ccsu.edu} \\
\vskip .2in
Institute of Cosmology\\
Department of Physics and Astronomy\\
Tufts University\\
Medford, Massachusetts 02155\\
\end{center}

\vspace*{1cm}
\begin{abstract}
There has been much recent work on quantum 
inequalities to constrain negative energy. These are 
uncertainty principle-type restrictions on the magnitude 
and duration of negative energy densities or fluxes. 
We consider several examples of 
apparent failures of the quantum inequalities,  
which involve passage of an observer through regions where the negative 
energy density becomes singular. We argue that this type of situation
requires one to formulate quantum inequalities using sampling functions
with compact support. We discuss such inequalities, and argue that they 
remain valid even in the presence of singular energy densities.    
\end{abstract}
\newpage

\baselineskip=14pt

\section{Introduction}
\label{sec:intro}
It has been known for some time that, unlike  
classical physics, quantum field theory allows the local energy 
density to be negative \cite{EGJ,Kuo}, and even unboundedly negative 
at a single spacetime point. These situations imply violation of 
the weak energy condition \cite{HE}: 
$T_{\mu \nu} u^{\mu} u^{\nu} \geq 0$, 
for all causal vectors $u^{\mu}$. On the other hand, if field theory 
places no constraints on negative energy, then it might 
be possible to produce gross macroscopic effects. Such 
effects might include: violation of the second law of 
thermodynamics \cite{F78,D82}, violation of the cosmic censorship 
hypothesis \cite{FR90,FR92}, traversable wormholes \cite{MT}, 
warp drives \cite{A,KRAS}, and time machines 
\cite{MTY,E,HCP}, to name a few. 
As a result, there has been much activity in recent years to 
determine what constraints, if any, quantum field theory places on 
negative energy. 

One approach involves averaging the local energy conditions 
over timelike or null geodesics. (See Refs. \cite{FW,FRBH} for 
discussion and references.) Another approach 
\cite{F78,F91,FR95,FR97} entails 
multiplying the renormalized expectation value of the energy 
density (or flux) by a sampling function, i.e., a peaked function 
of time whose time integral is unity. One convenient choice is the 
Lorentzian function peaked around $\tau=0$,
\begin{equation}
q(\tau) = \frac{\tau_0}{[\pi ({\tau}^2+{\tau_0}^2)]} \, ,  \label{eq:sampfnt}
\end{equation}
 where 
$\tau_0$ is the characteristic width of the sampling function, i.e., 
the ``sampling time''. 

Let $T_{\mu\nu}$ be the renormalized expectation value
of the stress tensor taken in an arbitrary quantum state $|\psi\rangle$.
Then $T_{\mu\nu}u^\mu u^\nu$ is the local energy density measured by an observer
with four-velocity $u^\mu$. For a quantized massless, 
minimally coupled scalar field in
four-dimensional Minkowski spacetime, the following 
inequality has been derived \cite{FR95,FR97} for timelike 
geodesic observers: 
\begin{equation}
\hat \rho = {{\tau_0} \over \pi}\, \int_{-\infty}^{\infty}\,
{{T_{\mu\nu} u^{\mu} u^{\nu}\, d\tau}
\over {{\tau}^2+{\tau_0}^2}} \geq
-{3\over {32 {\pi}^2 {\tau_0}^4}}\,,  \label{eq:4DENQI}
\end{equation}
for all $\tau_0$, where $\tau$ is the observer's proper time. 
(Our units are taken to be $\hbar=G=c=1$.) 
Similar bounds have also been derived for the massive 
scalar and electromagnetic fields \cite{FR97}. 
These constraints, which have come to be known as 
``quantum inequalities'' (QIs), are uncertainty principle-like 
bounds which restrict the magnitude and duration of 
negative energy effects. However, it should be noted that 
the energy-time uncertainty principle was not used as input 
to derive the QIs; they arise directly from quantum field theory. 
More recently, QI bounds have been proven in static curved 
spacetimes as well \cite{PF971,Song,PFGQI}. For the massless scalar 
field in two-dimensional Minkowski spacetime, generalized 
QI bounds have been derived for arbitrary sampling 
functions \cite{FLAN}. 

The original QI bounds were derived for Minkowski 
spacetime, and shown to hold for all choices of sampling 
time $\tau_0$. It was argued in Ref. \cite{FRWH} that the  
flat spacetime QIs should also hold in a curved spacetime 
and/or one with boundaries, if one restricts the choice of sampling 
time to be much less than either the smallest local proper radius of 
curvature or the smallest proper distance to any boundaries. 
In particular, it was shown that in this limit the Casimir effect satisfies 
the QI bound. This argument essentially says 
that one does not have to know the large-scale curvature of the 
universe in order to use flat space quantum field theory to 
predict the outcome of a laboratory-scale experiment. 
If the QI is applied in general spacetimes in the short sampling time 
limit, then it was shown that the bound severely constrains the 
geometry of traversable wormholes. More specifically, either the 
wormhole must be no larger than a few thousand Planck lengths in 
size, or, if the wormhole is macroscopic, there must be large 
discrepancies in the length scales which characterize the wormhole 
geometry, e.g., the negative energy must be confined to a band 
around the throat which can be no thicker than a few thousand Planck 
lengths. It was argued in Ref. \cite{FRWH} that, on 
dimensional grounds, one would not expect nonlocal curvature 
terms to produce significant contributions to the renormalized 
energy density over macroscopic length scales unless one 
introduced large dimensionless coupling constants into the 
theory or enormous numbers (e.g., $\sim 10^{62}$) of fields. 
In this sense, the conclusions of Ref. \cite{FRWH} are not at 
odds with some recent claims \cite{Hetal,KHAT}. Similar analyses 
 apply the flat spacetime QI, in the short sampling time limit, to the 
``warp drive'' spacetime of Alcubierre \cite{A,PFWD} and to the 
``superluminal subway'' spacetime of Krasnikov \cite{KRAS,ER}, and 
arrive at even more stringent constraints on the physical 
realizability of these spacetimes. 
Strong evidence for the validity of the short 
sampling time approximation has been provided by recent 
analyses  \cite{Song,PFGQI}. 
These show that for any static observer (geodesic or not) in 
any static spacetime, the QI reduces to the Minkowski 
spacetime form, Eq.~(\ref{eq:4DENQI}), in the 
short sampling time limit.  

Krasnikov \cite{KRAS} has recently pointed out that, in certain 
circumstances, the QIs might fail even in the short sampling time limit. 
He cites the specific example of a massless scalar field in the conformal 
vacuum state in two-dimensional Misner spacetime. For any geodesic 
observer and {\it any} sampling time, $\tau_0$, he observes that 
$\hat\rho = -\infty$ on the Cauchy horizon, and that $\hat\rho$ 
diverges to 
$-\infty$ as the observer approaches the Cauchy horizon. 
Krasnikov concludes that the QIs do not hold in this situation, 
and argues that similar failures should occur in the case 
where one ``almost transforms'' a traversable 
wormhole into a time machine \cite{COM1}. 

In this paper we give other examples of apparent 
failures of the QIs, which arise when there are singular energy densities.
 We argue that the problem arises when one employs
a sampling function, such as the Lorentzian function, with an infinite
``tail''. If one formulates the quantum inequalities in terms of sampling
functions with compact support, then the relevant integrals are finite,
so long as one samples outside  the region where the energy density becomes 
singular. We further argue that the physical content of the quantum inequalities
as restrictions on the magnitude and extent of negative energy is essentially
the same as found in previous work.

\section{Sampling Functions and Divergent Energy Densities}
\label{sec:SFDED}
\subsection{A Representative Example: The Flat Plate}
\label{sec:plate}

Consider a minimally coupled scalar field in four-dimensional 
Minkowski spacetime with a single plane boundary, which is 
located at $z=0$. We assume that the field is in the vacuum state, 
and that an observer approaches the boundary at constant 
velocity along the $z$-axis. We take the observer's equation 
of motion to be 
\begin{equation}
z(\tau)= v \gamma (\tau - \tau_c) \,,
\label{eq:EQM-2D}
\end{equation}
where $\tau$ 
is the observer's proper time, $\tau_c$ is the proper time at 
which the observer collides with the plate, and 
$\gamma= 1/ \sqrt{1-v^2}$. The corresponding four-velocity is 
$u^{\mu}= \gamma (1,0,0,v)$. The renormalized expectation values 
of the stress-tensor components for the quantum field are given 
by \cite{FULLING} 
\begin{equation}
T_{tt} = -T_{xx} =-T_{yy} =- \frac{1}{16 \, {\pi}^2 \, z^4} \,,   \label{eq:PEN}
\end{equation}
and 
\begin{eqnarray}
T_{zz} &=& 0  \,.
\end{eqnarray}  
We see that the energy density diverges as $z^{-4}$. 
(Such a divergence does not occur for the massless 
conformally coupled scalar field or for the electromagnetic 
field, in the plane boundary case. However, divergences do 
occur in the case of curved boundaries in flat spacetime \cite{DC}.) 

The energy density in this observer's frame is 
\begin{equation}
T_{\mu \nu} u^{\mu} u^{\nu} = {\gamma}^2 \, T_{tt} = 
-\frac{1}{16 \, \pi^2 \, v^4 \, {\gamma}^2 \, {(\tau - \tau_c)}^4} \,.
\label{eq:4DPLATE-ED}
\end{equation}
If we insert this expression into Eq.~(\ref{eq:4DENQI}), we obtain 
\begin{equation}
\hat \rho = 
-\frac{\tau_0}{16 \, \pi^3 \, v^4 \, {\gamma}^2} \, \int_{-\infty}^{\infty} \, 
\frac{d\tau}{ {(\tau - \tau_c)}^4 \, (\tau^2 + {\tau_0}^2) } \, .
\label{eq:PQI}
\end{equation}
This integral apparently diverges due to the singularity of the integrand 
as $\tau \rightarrow \tau_c$, corresponding to $z \rightarrow 0$. This is
independent of the choices both  of  $\tau_0$ and of  $\tau_c$.  

   From this example, one can see that the apparent failure 
of the QI occurs because the tail of the sampling function 
intersects the region of singular negative energy density. 
Furthermore, this problem may arise in more general cases 
for sampling functions which do not have compact support. 
For any sampling function which has a tail, one can always 
construct a quantum state designed so that  
the temporal asymptotic growth of the magnitude of the negative 
energy density overcomes any falloff of the chosen sampling function. 
However, the original choice of the 
Lorentzian sampling function, Eq.~(\ref{eq:sampfnt}), was made simply for
 mathematical convenience. In this paper we will argue that the problem posed 
above can be remedied by using compactly-supported sampling 
functions, that is, functions which are identically zero outside a finite
interval.

First, it is of interest to note that in some cases an alternative solution is
available. Although the integral in Eq.~(\ref{eq:PQI}) is apparently divergent,
it can in fact be defined as a ``generalized principal value '' integral 
 \cite{Davies}. The basic idea is to perform successive
integrations by parts. Consider the integral
\begin{equation}
I = \int_{-\infty}^{\infty} \, \frac{f(\tau)}{ {(\tau - \tau_c)}^4} \; d\tau \,,
\end{equation}
where $f(\tau)$ and its first three derivatives are finite everywhere, including
as $|\tau| \rightarrow \infty$. If we perform three successive integrations
by parts, the boundary terms all vanish and the result is
\begin{equation}
I = \frac{1}{6}\,\int_{-\infty}^{\infty} \, 
                     \frac{f^{'''}(\tau)}{ {\tau - \tau_c}} \; d\tau \,,
\end{equation}
where the remaining integral may be defined as a conventional principal value.
For the integral in Eq.~(\ref{eq:PQI}), $f(\tau) = (\tau^2 + \tau_0^2)^{-1}$,
and we find (with the aid of the symbolic algebra routine MACSYMA)
\begin{equation}
I = \pi\, \frac{(\tau_c^2-2\tau_0\tau_c-\tau_0^2)
(\tau_c^2+2\tau_0\tau_c-\tau_0^2)}{\tau_0 (\tau_c^2 + \tau_0^2)^4} \,.
\end{equation}
Equation (\ref{eq:PQI}) now becomes
\begin{equation}
\hat \rho = -\frac{(b^2-2b-1)(b^2+2b-1)}
{16 \, \pi^2 \, v^4 \, {\gamma}^2\, \tau_0^4\, (b^2+1)^4} \, ,
\end{equation}
where $b= \tau_c/\tau_0$. We want to choose the sampling time to be small
compared to the proper spatial distance to the plate, 
$\tau_0 \ll v \tau_c < \tau_c$. In this limit,
\begin{equation}
\hat \rho \approx -\frac{1}{16 \, \pi^2 \,  v^4 \,{\gamma}^2\, \tau_c^4} 
           \gg -\frac{1}{16 \, \pi^2 \, \tau_0^4} \,.
\end{equation}
Thus a quantum inequality of the form of Eq.~(\ref{eq:4DENQI}) is in fact
satisfied. This method can be used whenever the observer passes through
an energy density which diverges symmetrically as an inverse integral power
of proper time on either side of a boundary. It would not work, for example,
if the observer were to stop abruptly at $z=0$.

\subsection{A Two-Dimensional QI with a Compactly-Supported Sampling 
Function}

Consider the sampling function given by 
\begin{equation}
f(\tau)  = \left\{\matrix{0  \,, &  \,\, \tau < - \tau_0/2 \cr
(1/ \tau_0) \, 
[1+ {\rm cos}(2 \pi \tau / \tau_0) ] \,,
& \, -\tau_0/2 \leq \tau \leq \tau_0/2 \cr
0 \,, &  \,\, \tau > \tau_0/2}\right. \,.
\label{eq:CSSF}
\end{equation}
Flanagan has shown \cite{FLAN} that for a massless scalar 
field in two-dimensional Minkowski spacetime 
\begin{equation}
\hat\rho \geq -(1/{24 \pi}) \, \int_{-\infty}^{\infty} \, d\tau \,
\frac{{[g'(\tau)]}^2}{g(\tau)} \,,
\label{eq:FLANQI}
\end{equation}
where $g(\tau)$ is an arbitrary sampling function. 
If we substitute the sampling function given by 
Eq.~(\ref{eq:CSSF}) into Eq.~(\ref{eq:FLANQI}), we 
obtain the following QI:
\begin{equation}
\hat\rho \geq -\frac{\pi}{6 {\tau_0}^2} \,.
\label{eq:2DQI-CS}
\end{equation}
Although a similar inequality using a compactly-supported 
sampling function in four-dimensional Minkowski spacetime 
is not yet in hand, it is quite plausible to conjecture that such 
a QI exists. We now turn to showing that compactly-supported sampling 
functions may be used to resolve the apparent difficulties posed by 
divergent energy densities.

\section{Applications of QIs with Compactly-Supported Sampling Functions}
\label{sec:Appl}
\subsection{Flat Spacetime with Boundaries}
\label{sec:Fst-b}
\subsubsection{The Two-Dimensional Plate}
\label{sec:2dplate}
In a two-dimensional flat spacetime, unlike the case of 
four dimensions, the energy density for a minimally 
coupled scalar field does not diverge at a boundary under the 
imposition of Dirichlet boundary conditions. However, a non-minimally coupled
field does have such a divergence. The stress tensor is given by 
\begin{eqnarray}
T_{tt} &=& - \frac{A}{z^2} \,, \\ 
T_{zz} &=& 0 \,, 
\label{eq:2DP-ST}
\end{eqnarray} 
where $A = -\xi/2\pi$, and $\xi$ is the conformal coupling parameter. Here we 
assume that $A > 0$ \quad ($\xi < 0$) and of order one. Consider a geodesic 
observer whose equation of motion is given by 
Eq.~(\ref{eq:EQM-2D}). The energy density in this observer's 
frame is given by 
\begin{equation}
\rho(\tau) = T_{\mu \nu} u^{\mu} u^{\nu} =  
-\frac{A}{v^2  \, {(\tau - \tau_c)}^2} \,.  \label{eq:rho_2d}
\end{equation}
If we fold this into the compactly-supported sampling function 
given by Eq.~(\ref{eq:CSSF}), we obtain 
\begin{equation}
\hat\rho = -\frac{A}{v^2 \,\tau_0} \, \int_{-\tau_0/2}^{\tau_0/2} \,
\frac{d\tau}{ {(\tau - \tau_c)}^2} \, 
[1 + {\rm cos}(2 \pi \tau / \tau_0)] \,.  
\end{equation}
However, since the energy density is 
decreasing monotonically, as shown 
in Fig. 1, then $\hat \rho$ must be greater than the energy 
density at the end of the sampling interval, i.e., 
\begin{equation}
\hat\rho > \rho(\tau_0/2) \,.  \label{eq:rhoend}
\end{equation}
We can now ask under what circumstances will it be true that
\begin{equation}
\rho(\tau_0/2) \geq 
- \frac{\pi}{6{\tau_0}^2} \,.
\label{eq:QIA}
\end{equation}
Using the fact that the proper distance to  $z=0$,
as measured from the end of the sampling interval, 
is given by 
\begin{equation}
\ell = v \, [\tau_c - \tau_0/2] \,,
\end{equation}
and that in the present case
\begin{equation}
\rho(\tau_0/2) = -\frac{A}{v^2  \, {[\tau_c - \tau_0/2]}^2}
               = -\frac{A}{\ell^2} \,,
\end{equation}
we see that Eq.~(\ref{eq:QIA}) will be satisfied when 
\begin{equation}
\ell \geq {\biggl(\frac{6 A}{\pi } \biggr)}^{1/2} \, \tau_0 \,.
\end{equation}
Hence we see that our two-dimensional flat spacetime QI,  
Eq.~(\ref{eq:2DQI-CS}), will be satisfied when 
\begin{equation}
\tau_0 \ll \ell \,,
\end{equation}
that is, when the observer's sampling time is much 
smaller than the proper distance to the boundary.

\begin{figure}
\begin{center}
\leavevmode\epsfysize=10cm\epsffile{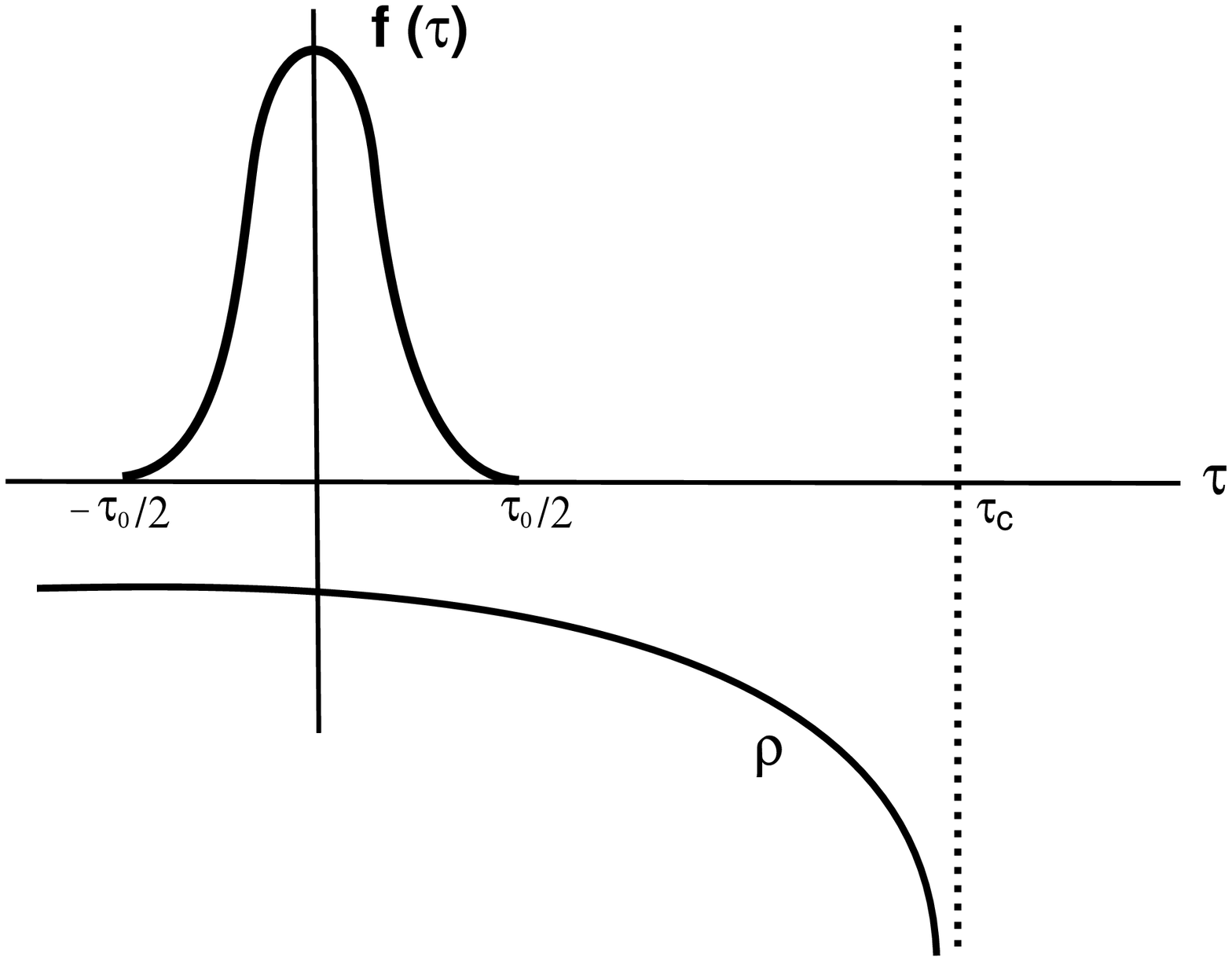}
\label{Figure 1}
\end{center}
\begin{caption}[]

The compactly-supported sampling 
function $f(\tau)$, and the 
energy density $\rho$, seen by an ingoing geodesic observer, 
plotted as a function of the proper time, $\tau$. The width of the 
sampling function, i.e., the sampling time, is $\tau_0$. The 
energy density seen by the observer diverges negatively as 
$\tau \rightarrow \tau_c$.
\end{caption}
\end{figure}

\subsubsection{The Four-Dimensional Plate}
\label{sec:4dplate-rev}
Although a QI which uses a compactly-supported 
sampling function has not been derived in 
four-dimensional flat spacetime as yet, 
it is reasonable to assume 
that an analog of Eq.~(\ref{eq:2DQI-CS}) exists. On 
dimensional grounds, such a QI for a massless 
scalar field should have the form 
\begin{equation}
\hat \rho \geq - \frac{\alpha}{ {\tau_0}^4} \,, \label{eq:4DQI}
\end{equation}
where $\alpha$ is a positive constant. 
For the purposes of the following argument we 
shall assume that such a QI exists, with $\alpha$ of the order of or less 
than unity. 

Let us return to the case of the observer approaching 
a flat plate in four-dimensional flat spacetime, discussed 
in Sec.~\ref{sec:plate}, and fold Eq.~(\ref{eq:4DPLATE-ED}) 
into the compactly-supported sampling function given 
by Eq.~(\ref{eq:CSSF}). The energy density in the 
observer's frame now decreases as $-1/{(\tau - \tau_c)}^4$. 
Repeating the argument for the two-dimensional case, we 
find that our conjectured QI will hold if
\begin{equation}
\hat \rho \geq  \rho(\tau_0/2)  
    \geq - \frac{\alpha}{ {\tau_0}^4} \, .     \label{eq:QIB}
\end{equation}
  From Eq.~(\ref{eq:4DPLATE-ED}), we find that this will be true if
\begin{equation}
\tau_0 \leq {(16 \pi^2 \alpha \, \gamma^2 )}^{1/4} \, \ell \,, 
\label{eq:TAU1}
\end{equation}  
where $\ell$ is again the proper distance to the plate. 
(Unlike the two-dimensional case, the 
factors of $\gamma$ do not cancel out.) When 
$v \rightarrow 0$, $\gamma \rightarrow 1$, and 
therefore the QI is satisfied when $\tau_0 \ll l$. 
As $\gamma$ gets larger, the condition 
Eq.~(\ref{eq:TAU1}) becomes easier to satisfy. 

Note that in this and the previous subsection, we have defined the proper
distance to the plate to be measured from the end of the sampling interval.
However, if $\tau_0 \ll \ell$, this is approximately the distance from the 
middle of the sampling interval. In subsequent subsections, we shall use
the latter definition.

\subsection{The Boulware Vacuum at $r=2M$}
\label{sec:Boulware}
\subsubsection{Two Dimensions}
\label{sec:2D}
The two-velocity of an ingoing geodesic observer in 
two-dimensional Schwarzschild spacetime is given by 
\begin{equation}
u^\mu = (u^t, u^r) = \Bigl({{dt} \over {d\tau}}, {dr\over {d\tau}} \Bigr) =
        \Bigl( {k \over C},  - \sqrt{k^2 -C} \Bigr).  \,
\label{eq:geod}
\end{equation}
where $C = 1 - 2M/r$ and $k$ is the energy per unit rest mass of the observer.
   From Eq.~(33) of Ref. \cite{FRBH}, we have that 
\begin{equation}
T_{\mu\nu} {u^\mu} {u^\nu}= {1\over {24\pi}}\,C^{-2}\, \Biggl\{
k^2\,\Biggl[ {{6M^2} \over{r^4}} - {4M\over {r^3}} \Biggr]\,
+\,{{CM^2} \over {r^4}} \Biggr\}.                  
\label{eq:brho2d}
\end{equation}
Note that this quantity is negative everywhere for 
$r \geq 2M$ \cite{INOUT}, and diverges at $r=2M$.

   From $r=\infty$ to $r=2M$, $C$ varies from $1$ to $0$. Consider 
the ultrarelativistic limit, $k \gg 1$, corresponding to an observer 
shot inward at high velocity. Then from Eq.~(\ref{eq:geod}), 
we have, to first order in $k^{-1}$,
\begin{equation}
\tau \sim \tau_c - \frac{rC}{k} \,,
\label{eq:tau}
\end{equation}
where $\tau_c$ is the proper time at which the observer 
reaches $r=2M$. As the observer approaches the horizon, the local energy
density varies as
\begin{equation}
T_{\mu\nu} {u^\mu} {u^\nu} \sim - \frac{1}{48 \pi (\tau_c -\tau)^2} \,.
          \label{eq:rho2Dhorizon}
\end{equation}
We can think of the horizon in the Boulware vacuum as a singular boundary
analogous to the flat space examples discussed in Sect.~\ref{sec:Fst-b}. 
Note that in the infalling observer's rest frame near the horizon, the $r=2M$ 
boundary is approaching at nearly the speed of light. As the horizon is
approached, the proper distance in this observer's frame to the boundary
 from the point $\tau=0$ (which is always the midpoint of our sampling 
interval)  is approximately $\ell = \tau_c$.  If we select a 
sampling time $\tau_0 \ll \ell$,  the sampling function is zero at the
horizon and the flat space
form of the quantum inequality, Eq.~(\ref{eq:2DQI-CS}), is satisfied, just as in the 
example in Sect.~\ref{sec:2dplate}.

\subsubsection{Four Dimensions}
\label{sec:4D}
Visser has recently given an approximate analytic expression for 
the renormalized stress-tensor components of a conformally 
coupled scalar field in the Boulware vacuum state in 
four-dimensional Schwarzschild spacetime \cite{VBOUL}. 
Our original QI was proven for the minimally 
coupled, rather than the conformally coupled scalar field. In 
light of the recent proofs of 
similar QIs for the massive scalar field and the electromagnetic 
field \cite{FR97,PF971}, it seems highly likely that such a bound 
should also hold for the conformally coupled scalar field 
as well. For the sake of the following argument, 
we will assume this to be true. 

If Visser's Eqs. (8) and (9) are transformed from the static 
orthonormal frame back into the usual Schwarzschild $t,r$ 
coordinates, one obtains  
\begin{eqnarray}
T_{tt} &=& -3 \,p_{\infty} \, x^6 \, 
\frac{[40-72 x+33 x^2]}{(1- x)} \\
\label{eq:Ttt4d}
T_{rr} &=& \,p_{\infty} \, x^6 \, 
\frac{[8-24 x+15 x^2]}{ {(1- x)}^3} \,,
\label{eq:Trr4d}
\end{eqnarray}
where $x \equiv 2M/r$, and 
\begin{equation}
p_{\infty} = \frac{1}{90{(16 \pi)}^2{(2M)}^4} \,.
\label{eq:pinfdef}
\end{equation}
For an infalling geodesic observer with 
$u^\mu = ( k / (1-x),  - \sqrt{k^2 - (1- x)}, 0, 0)$, we have that 
\begin{equation}
T_{\mu \nu} u^{\mu} u^{\nu} =
\frac{- p_{\infty} \, x^6 \, 
(15 x^3 - 84 k^2 x^2 - 39 x^2 + 192 k^2 x + 32 x -112 k^2 -8)}
{ {(x-1)}^3} \,.
\label{Tmunu4d}
\end{equation} 
If we take $k \gg 1$, and use Eq.~(\ref{eq:tau}), we can express the energy 
density near the horizon as 
\begin{equation}
T_{\mu \nu} u^{\mu} u^{\nu} \approx
- \frac{32 M^3 p_{\infty}}{k (\tau_c -\tau)^3} \,.
\end{equation}
As discussed in the previous subsection, near the horizon,
the observer's proper distance to $r=2M$ is $\ell \approx  \tau_c$.
If $\tau_0 \ll \ell$, then the quantum inequality, Eq.~(\ref{eq:4DQI}),
will be satisfied if
\begin{equation}
\tau_0 < [45(16 \pi)^2 k \,\alpha\, M \,\ell^3]^{\frac{1}{4}} \,.
\end{equation}
However, this will in fact be the case because $k \gg 1$ and $\tau_0 \ll
\ell < M$.

\subsection{The Singularity at $r=0$ in Black Hole Spacetimes}
 \label{sec:origin}

 Perhaps the most serious example of a singular energy density arises at 
$r=0$, the curvature singularity of a black hole. Unlike the other examples 
discussed in this paper, this singular energy density cannot be explained away
as being due to an unphysical choice of quantum state, as is the case of the
Boulware vacuum at the horizon, or an unphysical boundary condition, as in 
the case of the perfectly reflecting plate in Sect.~\ref{sec:4dplate-rev}. 
This singular energy density is essentially independent of the quantum state.
In this case, the generalized principal value method discussed in 
Sect.~\ref{sec:plate} cannot be utilized, as the observer cannot pass beyond
$r=0$. However, the use of compactly-supported sampling functions is still
successful. We will restrict our attention
to the case of a two-dimensional black hole, as the form of the stress tensor
near $r=0$ is not known in the four-dimensional case. From 
Eq.~(\ref{eq:brho2d}),
or from the corresponding expression for the Unruh vacuum state, one finds 
that near the origin of a two-dimensional black hole,
\begin{equation}
T_{\mu\nu} {u^\mu} {u^\nu} \sim - \frac{M}{48 \pi r^3} \,.
\end{equation}
The geodesic equation, Eq.~(\ref{eq:geod}), implies that for small $r$
\begin{equation}
r(\tau) \approx 
\left[\frac{3 \sqrt{2M}}{2} (\tau_c -\tau) \right]^{\frac{2}{3}} \, ,
\end{equation}
where $\tau_c$ is again the proper time at which the singularity is reached.
The energy density can be expressed in terms of the proper time as
\begin{equation}
T_{\mu\nu} {u^\mu} {u^\nu} \sim - \frac{1}{216 \pi (\tau_c -\tau)^2} \,.
\end{equation}
Note that this result has the same form as Eqs.~(\ref{eq:rho_2d})
and (\ref{eq:rho2Dhorizon}). A two-dimensional spacetime is characterized by
a single component of the Riemann tensor, which we may take to be the scalar
curvature. In the case of 2D Schwarzschild spacetime, this is
\begin{equation}
R = \frac{4M}{r^3} \,.
\end{equation}
Define the proper local radius of curvature by
\begin{equation}
r_c = \frac{1}{\sqrt{R}} \sim \frac{3 \sqrt{2}}{4} (\tau_c -\tau)\,  \qquad
                   {\rm as}  \quad    \tau \rightarrow \tau_c \, .
\end{equation}
We see that near $r=0$, this local radius of curvature and the proper time
for an observer to reach the singularity, $\tau_c -\tau$, are proportional
to one another. Thus if we require that the sampling time satisfy 
$\tau_0 \ll r_c$, we again find that the flat space form of the quantum
inequality, Eq.~(\ref{eq:2DQI-CS}), is satisfied.

\subsection{Misner Space}
\label{sec:misner}

A further example of a singular energy density arises in Misner space. The
two-dimensional version of this example was cited by Krasnikov as the possible
counterexample to quantum inequalities based upon noncompactly-supported sampling 
functions. The stress tensor in this two-dimensional version has recently
been discussed in detail by Cramer and Kay \cite{CK97}.
Hiscock and Konkowski \cite{HK82} have calculated the quantum stress tensor
for a massless conformally coupled scalar field in the four-dimensional version
of Misner space, so we may use their results to demonstrate that quantum
inequalities using compactly supported sampling functions are meaningful 
in this space.
Misner space is a locally flat spacetime with periodic identifications. It may
be represented by the metric
\begin{equation}
ds^2 = -dt^2 +t^2 dx^2 +dy^2 +dz^2 \, ,  \label{eq:misner}
\end{equation}
with the points $(t,x,y,z)$ and $(t,x+na,y,z)$
identified with one another, where $n$ is any integer and $a$ is a positive
constant. Misner space is a portion of Minkowski space, and the metric may
be transformed to the Minkowski form
\begin{equation}
ds^2 = -dy_0^2 + dy_1^2 + dy_2^2 + dy_3^2
\end{equation}
by means of the transformation
\begin{equation}
y_0 = t \cosh x, \quad y_1 = t \sinh x, \quad y_2 =y, \quad y_3 =z \,.
\end{equation}

The quantum stress tensor is divergent everywhere on the Cauchy horizon at
$t=0$, including the  ``quasiregular singularity'' at $y_0 = y_1$.  
(See Figure 2.) Hiscock and Konkowski show that the expectation value of the
stress tensor in the conformal vacuum state is given in the coordinates of
Eq.~(\ref{eq:misner}) by
\begin{equation}
T_{\mu\nu} = \frac{K}{12 \pi^2 t^4} \, {\rm diag}(-1, -3t^2, 1, 1) \, ,
\end{equation}
where
\begin{equation}
K = \sum_{n=1}^\infty \; \frac{2 + \cosh(na)}{[\cosh(na) - 1]^2} \, .
                                          \label{eq:defK}
\end{equation}
They further show that in the frame of a geodesic observer, the energy density
diverges on the Cauchy horizon as $\tau^{-3}$, where $\tau$ is proper time
measured from the horizon. Note that here we are discussing a situation where
the divergent energy density is in the observer's past, as illustrated in
Figure 2. However, one could equally well discuss the time reversed situation
where the singular energy is encountered in the future. In the special case
in which a geodesic observer meets the  quasiregular singularity, the energy
density diverges as $\tau^{-4}$. As this case seems to pose the strongest 
challenge for quantum inequalities, we will focus our attention here. 

\begin{figure}
\begin{center}
\leavevmode\epsfysize=10cm\epsffile{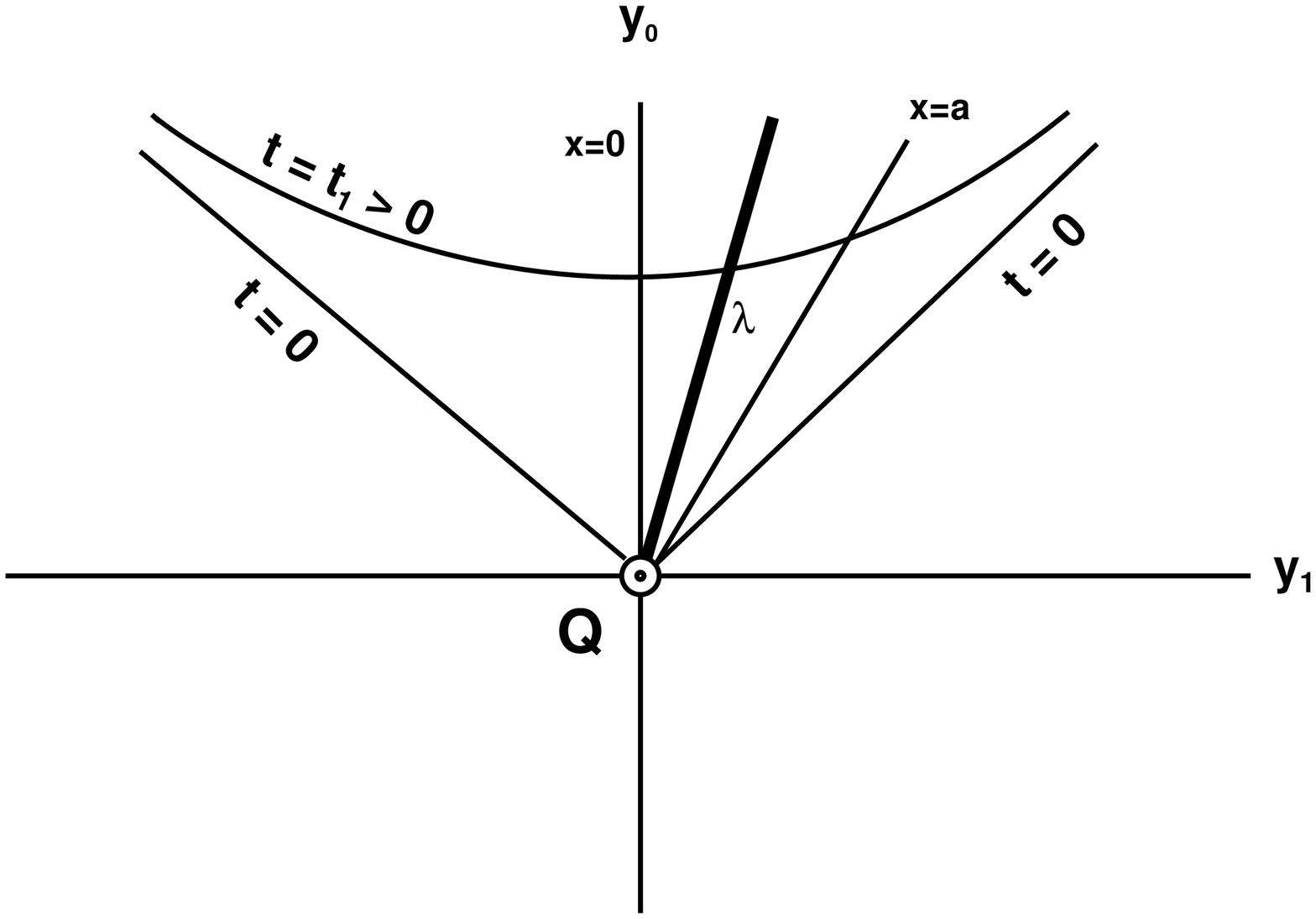}
\label{Figure 2}
\end{center}
\begin{caption}[]

A two-dimensional section of (four-dimensional) 
Misner spacetime, shown as the upper quadrant of 
Minkowski spacetime. The coordinates $(y_0, y_1)$ are 
Minkowski coordinates, whereas $(t, x)$ are Misner space 
coordinates. The straight lines with slope 
less than $45$ degrees are lines of constant $x$. The 
lines $x=0$ and $x=a$ are identified with one another. 
The line labeled $\lambda$ is the worldline of a 
geodesic observer who passes through the 
quasi-regular singularity, {\bf Q}. The Cauchy horizon is 
represented by the $t=0$ lines. (Note that in this 
representation the singularity is in the past.) The curve 
labeled $t = t_1 > 0$ is an arbitrary $t = \, const$ curve.
\end{caption}
\end{figure}

The observer in question moves along the path
\begin{equation}
y_1 = v_x \,y_0, \quad y_2 = v_y \,y_0, \quad y_3 = v_z \,y_0 \, .
\end{equation}
In the coordinates of Eq.~(\ref{eq:misner}),  $v_x = \tanh x$, and hence
the observer's path is a line of constant $x$. The components of the observer's 
four-velocity in these coordinates are
\begin{equation}
u^t = \frac{\gamma}{\cosh x}, \quad u^x = 0, \quad u^y = \gamma v_y,
\quad u^z = \gamma v_z \, ,   \label{eq:fourvel}
\end{equation} 
where, as usual, $\gamma = 1/\sqrt{1-(v_x^2 + v_y^2 + v_z^2)}$. (Note that
Hiscock and Konkowski use an unconventional definition of $\gamma$.)
The local energy density in this observer's frame near the singularity is
\begin{equation}
\rho = T_{\mu \nu} u^{\mu} u^{\nu} =
 - \frac{K}{12 \pi^2 \gamma^4 (1-v_x^2)^2\, \tau^4} \,. \label{eq:rho_misner}
\end{equation}
Note that
\begin{equation}
\frac{1}{\gamma^4 (1-v_x^2)^2} = 
\left(\frac{1- v_x^2 - v_y^2 - v_z^2}{1-v_x^2}\right)^2 \leq 1 \, ,
\end{equation}
so that unless $K \gg 1$, the quantum inequality Eq.~(\ref{eq:4DQI}) is
satisfied whenever the sampling time is chosen so that $\tau_0 \ll \tau$.

The latter condition is certainly  necessary in order that the spacetime
be Minkowskian over the time of the sampling, but it is by no means sufficient.
Space is compact in the $x$-direction, with a proper periodicity length which
goes to zero near $t=0$. We  see this from the following considerations.
The element of proper length in the $x$-direction is $d\ell = t \,dx$, and from
the first relation in Eq.~(\ref{eq:fourvel}), we have that 
$t = \tau \gamma/ \cosh x$. Thus, the proper periodicity length is
\begin{equation}
\ell = \gamma \tau \int_x^{x+a} \frac{dx}{\cosh x} = 
2 \gamma \tau \left[\tan^{-1} \left(e^{x+a}\right) - 
\tan^{-1} \left(e^{x}\right)\right] \,.  \label{eq:ell}
\end{equation}
In the limit of small $a$, this may be expressed as
\begin{equation}
\ell \approx \gamma \tau \frac{a}{\cosh x}  \,.  \label{eq:ell2}
\end{equation}
Spacetime is Minkowskian only on scales small compared to this length,
so we must also require that $\tau_0 \ll \ell$. The argument in the previous 
paragraph works except when $K \gg 1$. However, large $K$ arises only when
$a$ is small, which is precisely when $\ell \ll \tau$. From Eq.~(\ref{eq:defK})
we see that when $a \ll 1$,
\begin{equation}
K \approx \sum_{n=1}^\infty \; \frac{12}{a^4 n^4} \, = 
\frac{12 \zeta(4)}{a^4} \,.
                                          \label{eq:Ksmalla}
\end{equation}
The local energy density in the observer's frame,  Eq.~(\ref{eq:rho_misner}),
can be expressed as
\begin{equation}
\rho =  - \frac{K \cosh^4 x}{12 \pi^2 \gamma^4 \tau^4}\, .
                                                 \label{eq:rho_misner2}
\end{equation}
Thus the quantum inequality Eq.~(\ref{eq:QIB}) is satisfied if
\begin{equation}
\tau_0 < \frac{\gamma (12 \pi^2 \,\alpha\, K^{-1})^{1/4}}{\cosh x}\: \tau \,. 
                                           \label{eq:tau0misner}
\end{equation}
Equations (\ref{eq:Ksmalla}) and (\ref{eq:ell2}), and the relation $\zeta(4) =
\pi^4/90$, imply that for $a \ll 1$, Eq.~(\ref{eq:tau0misner}) becomes
\begin{equation}
\tau_0 < \left(\frac{90 \alpha}{\pi^2}\right)^{1/4}\, \ell  \,,
\end{equation}
which is indeed satisfied if $\tau_0 \ll \ell$. This confirms that the
quantum inequality holds in Misner space.

\section{Discussion}
\label{sec:disc}

In the preceeding sections, we have seen that various examples of singular
energy densities obey quantum inequalities, provided that these inequalities
are formulated using sampling functions with compact support. In order to 
 sample the region around a singular energy density, it is 
desirable that the sampling function be identically zero at the singularity.
Sampling functions with infinite tails, such as the Lorentzian function
Eq.~(\ref{eq:sampfnt}), lead to divergent integrals because the tail encounters
the  singularity. Clearly, this behavior is not realistic. In the case of a 
particle falling into a black hole, for example, one will get a divergent
integral regardless of where on the worldline one samples. A more reasonable 
outcome would be that the result of sampling while the particle is still
very far away from the black hole is independent of the future fate of the
particle. Quantum inequalities based upon compactly supported sampling functions
achieve this outcome. The generalized principal value method discussed in
Sect. \ref{sec:plate} is capable of rendering the integrals
associated with noncompactly supported sampling functions finite in some
cases, but not in the case of the singularity at $r=0$ in a black hole.

In any case, one does not expect truly divergent stress-energies to 
occur in reality. Various effects would be expected to smear out the 
divergences in physically realizable cases. For example, in the flat 
plate case, one expects the perfectly reflecting boundary condition 
imposed on the quantized electromagnetic field to break down for wavelengths 
smaller than about $\lambda_p=1/f_p$, where $f_p$ is the plasma frequency. 
Similarly, one would not expect the Boulware vacuum at $r=2M$ to be 
physically realizable, since the divergent stress-energy would 
produce a large backreaction which would presumably drastically alter 
the spacetime. A related argument could be made for the Misner and 
``almost time-machine'' wormhole spacetimes. These cases  
are all pathological in the sense that the quantum states 
do not have the Hadamard form on the horizon \cite{WQFTCST, KRW}. 

One can understand the origin of singular energy densities, such as 
Eq.~(\ref{eq:PEN}), on a surface on which the quantum scalar field, $\varphi$,
 satisfies vanishing boundary conditions as follows: $\varphi$ and its
time derivative, $\dot \varphi$, are conjugate variables. Hence they satisfy an 
uncertainty relation such that if $\varphi$ is precisely determined, then
$\dot \varphi$ must be completely uncertain. This means that 
$\langle \dot \varphi^2 \rangle$ and hence $T_{tt}$ must diverge. This
situation is analogous to that of a position eigenstate in single particle
quantum mechanics; such a state would have to have a completely uncertain
momentum, and hence an infinite mean energy. This suggests that the singular
energy density may disappear if the boundary's position is uncertain. This
has recently been proven to be the case for the flat plate example of 
Sec.~\ref{sec:plate}. In Ref. \cite{FS97} it is shown that if the plate is
in a quantum state where the position has a Gaussian probability distribution
of finite width, then the mean energy density is finite everywhere and 
approaches Eq.~(\ref{eq:PEN}) in the limit that this width vanishes. 

So long as the energy density is bounded below, one expects that even 
quantum inequalities based upon noncompact sampling functions such as 
Eq.~(\ref{eq:4DENQI}) to hold, provided that the sampling time is sufficiently
short. Here sufficiently short presumably means 
 $\tau_0 \ll \ell$, where $ \ell \sim {(\rho_{\rm max})}^{-1/4}$ 
and $\rho_{\rm max}$ is the maximum magnitude of the negative 
energy density.

\vskip 0.2 in
\centerline{\bf Acknowledgements}
The authors would like to thank Adam Helfer and Matt Visser 
for stimulating comments. TAR would like to thank the 
members of the Tufts Institute of Cosmology for their continuing 
hospitality while this work was being done. This research 
was supported by NSF Grant No. Phy-9507351, the John F. Burlingame Physics 
Fellowship Fund, and a CCSU/AAUP faculty research grant.


\begin{thebibliography}{--}
\bibitem{COM1} However, presumably one 
would have to make and maintain the wormhole prior to converting 
it into a time machine. We would expect 
the short sampling time limit to be valid in the negative energy 
density region near the throat. Therefore, it would seem that one 
would still have to deal at some stage with the large length 
discrepancies in the wormhole geometry discussed earlier.
\bibitem{INOUT} Note that the expression for 
                          $T_{\mu \nu} u^{\mu} u^{\nu}$ is the same for both 
                          ingoing and outgoing geodesic observers, since 
                          $T_{tr}=0$ in the Boulware vacuum state.
\end{thebibliography}
\end{document}